\documentclass[12pt]{article}
\usepackage{amssymb}
\usepackage{graphicx}

\newcommand{\blank}{}
\renewcommand{\theequation}{\blank \arabic{equation}}

\newcounter{dummy}{}
\newcommand{\letters}{    \setcounter{dummy}{\value{equation}}
    \renewcommand{\thedummy}{\blank \arabic{dummy}}
    \renewcommand{\theequation}{\thedummy\alph{equation}}
    \refstepcounter{dummy}
    \setcounter{equation}{0} }
\newcommand{\noletters}{    \setcounter{equation}{\value{dummy}}
    \renewcommand{\theequation}{\blank\arabic{equation}}  }
\newenvironment{mathletters}{\letters}{\noletters}

\begin{document}

\title{Electronic spectrum and tunnelling properties of multi-wall carbon nanotubes}
\author{A.~A.~Abrikosov,~jr.$^{1}$, D.V.Livanov$^{2}$, A.A.Varlamov$^{3}$ \\
$^{1}${\small Institute of Theoretical and Experimental Physics,}\\
{\small \ B. Cheremushkinskaya street 25, 117218 Moscow, Russia}\\
{\small \ }$^{2}${\small \ Moscow State Institute for Steel and Alloys }\\
{\small \ (Technological University),}\\
{\small \ Leninsky prospect 4, 119049 Moscow, Russia}\\
{\small \ }$^{3}${\small COHERENTIA-INFM UdR''Tor Vergata'', }\\
{\small \ Via del Politecnico, 00133 Roma, Italy}}
\date{\today}
\maketitle

\begin{abstract}
We develop a general approach to calculations of the electron spectrum of
metallic multi-wall carbon nanotubes (MWNT) with arbitrary number of coaxial
layers. It is based on the model with singular attractive potential of
equidistant conductive cylinders. The knowledge of one-electron spectrum
allows to construct the corresponding Green function and then to calculate
the entropy and density of states for MWNT. We analyze the tunnelling
between the nanotube and normal metal electrode. The possibility of direct
determination of one-electron density of states by measurements of the
tunnelling conductivity at low temperatures is proved and the necessary
restrictions on temperature are formulated. We discuss briefly the
conflicting experimental observations of electronic properties of MWNT.
\end{abstract}

 \smallskip

\noindent \textsc{keywords:} \qquad Nanotube, electron spectrum, density of
states.

\hyphenation{pseu-do-par-ticle pseu-do-par-ticles}

\section{Introduction}

Recently unusual properties of carbon nanotubes attract much interest. These
mesoscopic systems have demonstrated a remarkable interplay of
dimensionality, interaction and disorder \cite{Dek99}. The well-pronounced
one-dimensional character of electron motion in single-wall carbon nanotubes
(SWNT) with strongly separated levels of angular quantization led to
attempts to describe their behavior in terms of the Luttinger liquid model
\cite{EG97,KBF97,E99,BCLR99}. As soon as metallic SWNT has two conducting
channels, \ the $1D$ Luttinger liquid model may apply to this case when the
interchannel coupling is negligible.

Experimentally, several contradictory observations of the electronic
properties of SWNT at low temperatures were made. Such an evidence of the
Luttinger liquid behavior as a power law divergence of resistance when
temperature fell down to $10$ $K$\cite{BCLR99}, was recently attributed to
Coulomb blockade in tunnel junctions \cite{KKFD03}. The measured with the
use of low-resistive metallic contact low-temperature resistance of
individual SWNT, \cite{KKFD03} does not manifest traces of Luttinger liquid
behavior.

The multiwall nanotubes (MWNT), composed of several concentric graphite
shells show properties which are clearly consistent with the
weak-localization features of diffusive transport in magnetoconductivity and
zero-bias anomaly in the tunnelling density of states \cite
{BSSB99,BFPF00,KBMR03}. The intertube coupling in MWNT was demonstrated to
be essential for interpretation of experimental data on electric transport
properties of MWNT \cite{FPWH98,SKTL00}. On the other hand, if electrods
contact only the outermost tube in MWNT, it behaves as an individual
cylinder with the diameter an order of magnitude larger than that of SWNT
and may manifest ballistic properties \cite{SBSS99}.

In the present paper we study electron states in MWNT. In this case,
contrary to SWNT, the inter-tube electron hopping eliminates the specific
one-dimensional properties and we may solve the corresponding spectral
problem within the standard quantum mechanical many-electron approach. In
order to describe the electron spectrum of MWNT one has to take into
consideration that the motion of electron around the circumference of
individual tube is quantized. Due to the periodic boundary conditions an
integer number of wavelengths must fit around the tube. Along the tube,
however, electronic states are not restricted and electrons can move
ballistically. Because of the circumferential quantization, the electron
states in the tube do not form a single wide energy band. Instead, they
split into a number of one-dimensional sub-bands, with band onsets at
different energies. In MWNT an additional effect of intertube interaction
appears. It results in further splitting of energy levels. The
characteristic distance between the levels is governed by the probability of
electron tunnelling between adjacent tubes.

We model the metallic MWNT as a set of equidistant coaxial cylinders with
attractive Kronig-Penney type potential along the radius and study the
electron states of such a system. We develop an universal formalism for
numerical evaluation of electron energy levels. It allows to calculate
energy level splitting due to intertube hopping for MWNT with any number of
coaxial tubes. Then we consider the filling of the obtained band structure
by electrons and calculate such electronic properties as the one-particle
density of states, electronic contribution to entropy and, finally, the
tunnelling conductivity of MWNT. We also prove the possibility of direct
determination of the electron spectrum in MWNT by means of the tunnelling
microscopy and formulate the corresponding criterion on experimental
temperatures.

\section{Electrons on coaxial nanotubes}

\subsection{The model and notation}

We shall consider the system of graphite nanotubes as a set of equidistant
coaxial conducting cylinders. The number of cylinders $N$ is arbitrary.
Electrons are held to the cylinders by a singular attractive potential. The
single electron Hamiltonian takes the form:
\begin{equation}
H=-\frac{\hbar ^{2}\nabla ^{2}}{2m}-\sum_{i=1}^{N}U\,\delta (r-r_{i}),\qquad
\mathrm{with}\qquad r_{i}=r_{1}+(i-1)a.  \label{hamiltonian}
\end{equation}
Here $N$ is the number of coaxial tubes and $r_{i}$ are their radii. The
radius of the inner tube is $r_{1}$ and the adjacent tubes are separated by
distance $a\leq r_{1}$. We shall assume that the potential is strong.
Quantitatively this means that $U\gg \frac{\hbar ^{2}}{ma^{2}}$.

The Schr\"{o}dinger equation for electron reads:
\begin{equation}
-\frac{\hbar ^{2}\nabla ^{2}}{2m}\,\Psi -\sum_{i=1}^{N}U\,\delta
(r-r_{i})\,\Psi =E\,\Psi .  \label{Schroedinger}
\end{equation}
We shall solve the problem in cylindrical coordinates with $z$-axis pointing
along the tube. Let us take the electron wave function in the form
\begin{equation}
\Psi (z,\,r,\,\phi )=\sum_{n=0}^{\infty }\psi _{n}(r)\,\exp \left( \frac{i}{%
\hbar }p_{z}z+in\phi \right) ,  \label{psi}
\end{equation}
The electron has momentum $p_{z}$ and moves freely along the tube. The
angular motion of the electron is characterized by magnetic quantum number $%
n $. The potential affects only the radial part of the wave function:
\begin{equation}
-\frac{\partial ^{2}\psi _{n}}{\partial r^{2}}-\frac{1}{r}\frac{\partial
\psi _{n}}{\partial r}+\frac{n^{2}}{r^{2}}\psi _{n}-\frac{2mU}{\hbar ^{2}}%
\,\sum_{i=1}^{N}\delta (r-r_{i})\,\psi _{n}=\frac{2m}{\hbar ^{2}}\,\left( E-%
\frac{p_{z}^{2}}{2m}\right) \psi _{n}.  \label{r-eqn.}
\end{equation}
Note that the electrons are localized on the tubes and the right hand side
of the equation is negative.

The potential energy may be characterized by the effective width of $\delta $%
-wells:
\begin{equation}
\lambda =\frac{\hbar }{\sqrt{mU}}.  \label{lambda}
\end{equation}
It may be seen from the corresponding one-dimensional problem that $\lambda $
controls the spread of wave functions out of the tubes. Large values of $U$
correspond to the case $\lambda \ll a,\,r$.

Another important quantity is the spectral parameter. It has the dimension
of inverse length:
\begin{equation}  \label{kappa}
\kappa = \frac{1}{\hbar} \sqrt{p_z^2 - 2m E}.
\end{equation}
The spectral parameter sets the radial scale for wave function and its
values determine the discrete component of energy spectrum.

Our goal is to find $\kappa $ as a function of $\lambda $ depending on the
system geometry. After the standard rescaling $x=\kappa r$ the equation for $%
\psi _{n}(x)$ takes the form:
\begin{equation}
x^{2}\psi _{n}^{\prime \prime }+x\psi _{n}^{\prime }-(x^{2}+n^{2})\psi _{n}=-%
\frac{2}{\kappa \lambda }\sum_{i=1}^{N}x_{i}^{2}\delta (x-x_{i})\,\psi
_{n},\qquad \mathrm{with}\qquad x_{i}=\kappa r_{i}.  \label{x-eqn}
\end{equation}
We shall describe square integrable solutions of this equation in the next
section.

\subsection{Matching conditions}

The potential on the right hand side of equation (\ref{x-eqn}) is zero
almost everywhere excluding the points $x_{i}=\kappa r_{i}$. Between the
tubes the equation is of the modified Bessel type. Hence on the intervals $%
x_{i}\leq x\leq x_{i+1}$ the wave function $\psi _{n}$ must be a
superposition of modified Bessel functions:
\begin{equation}
\psi _{n}(x)=A_{i}^{n}I_{n}(x)+B_{i}^{n}K_{n}(x),\qquad \mathrm{for}\qquad
x\in \lbrack x_{i},\,x_{i+1}],\qquad x_{0}=0.  \label{superposition}
\end{equation}
The $2N+2$ coefficients $A_{i}^{n}$, $B_{i}^{n}$, $i=0,\ldots \,N$, are
determined by matching conditions at singular points of potential (\ref
{continuity}, \ref{derivatives}), boundary conditions (\ref{boundary}) and
overall normalization of the wave function (\ref{normalization}).

The first $N$ relations between $A^n_{i-1}$, $B^n_{i-1}$ and $A^n_i$, $B^n_i$
on the next interval come from the continuity of solutions, $\psi_n(x_i-0)=
\psi_n(x_i + 0)$. Substituting (\ref{superposition}) for $\psi_n$ we get:
\begin{equation}  \label{continuity}
A^n_{i-1} I_n (x_i) + B^n_{i-1} K_n (x_i) = A^n_i I_n (x_i) + B^n_i K_n
(x_i), \qquad i=1,\ldots\, N.
\end{equation}

The second set of conditions is obtained from matching the derivatives $\psi
_{n}^{\prime }$. They change at singular points $x_{i}$ stepwise and $\psi
_{n}^{\prime \prime }$ behaves like $\delta $-function. Integrating the two
sides of equation (\ref{x-eqn}) over an infinitesimal interval $%
(x_{i}-\epsilon ,\,x_{i}+\epsilon )$ we obtain for $\epsilon \rightarrow 0$:
\begin{equation}
\lim_{\epsilon \rightarrow 0}\int_{x_{i}-\epsilon }^{x_{i}+\epsilon
}x^{2}\psi _{n}^{\prime \prime }\,dx=x_{i}^{2}\left. \psi _{n}^{\prime
}\right| _{x_{i}-0}^{x_{i}+0}=-\frac{2x_{i}^{2}}{\kappa \lambda }\psi
_{n}(x_{i})\qquad \mathrm{and}\qquad \left. \frac{\psi _{n}^{\prime }}{\psi
_{n}}\right| _{x_{i}-0}^{x_{i}+0}=-\frac{2}{\kappa \lambda }.
\label{gen-derivatives}
\end{equation}
This leads to $N$ more equations of the form:
\begin{equation}
\frac{(A_{i}^{n}-A_{i-1}^{n})I_{n}^{\prime
}(x_{i})+(B_{i}^{n}-B_{i-1}^{n})K_{n}^{\prime }(x_{i})}{%
A_{i}^{n}I_{n}(x_{i})+B_{i}^{n}K_{n}(x_{i})}=-\frac{2}{\kappa \lambda }%
,\qquad i=1,\ldots \,N,  \label{derivatives}
\end{equation}
where we have used (\ref{continuity}) in the denominator.

The requirement of regularity of the wave function at zero and infinity
results in two more relations. Remember that the modified Bessel function $%
I_n (x)$ is regular at $x = 0$ and grows at $x\rightarrow\infty$ whereas $%
K_n (x)$, on the contrary, is singular at $x = 0$ but goes to zero at
infinity. Hence one must set
\begin{equation}  \label{boundary}
A^n_N = B^n_0 = 0.
\end{equation}

It may be easily seen that the listed $2N+2$ equations are not independent.
Namely, rescaling all $A\rightarrow \alpha A$, $B\rightarrow \alpha B$ does
not break the equalities. The missing requirement that helps to fix the
values of $A$ and $B$ is the normalization condition
\begin{equation}  \label{normalization}
\int_0^\infty |\psi_n (x)|^2 \, x\, dx = \int_0^\infty |\psi_n (r)|^2 \, r\,
dr = 1.
\end{equation}

We shall show how to solve equations (\ref{continuity}, \ref{derivatives},
\ref{boundary}) in the next section.

\subsection{The recursions \label{recursion}}

The standard way to find the spectrum of the described system of homogenous
equations would be to study the dependence of its determinant on spectral
parameter. We propose a recursive procedure that allows to deduce the
spectral equation without calculating the determinant.

The wave functions of electrons in the system of $N$ coaxial nanotubes are
given by formula (\ref{superposition}) where $2N+2$ coefficients $A_{i}^{n}$
and $B_{i}^{n}$ may be found up to a factor from the following system of $%
2N+2$ homogenous linear equations (we summarize once more (\ref{continuity},
\ref{derivatives}, \ref{boundary})):

\begin{mathletters}
\label{system}
\begin{eqnarray}
(A_{i}^{n}-A_{i-1}^{n})I_{n}(x_{i})-(B_{i-1}^{n}-B_{i}^{n})K_{n}(x_{i}) &=&0,
\label{systema} \\
i &=&1,\ldots \,N;  \nonumber \\
(A_{i}^{n}-A_{i-1}^{n})I_{n}^{\prime
}(x_{i})+(B_{i}^{n}-B_{i-1}^{n})K_{n}^{\prime }(x_{i})+\frac{2}{\kappa
\lambda }(A_{i}^{n}I_{n}(x_{i})+B_{i}^{n}K_{n}(x_{i})) &=&0,  \label{systemb}
\\
i &=&1,\ldots \,N;  \nonumber \\
B_{0}^{n} &=&0;  \label{systemc} \\
A_{N}^{n} &=&0  \label{systemd}
\end{eqnarray}
\end{mathletters}

This system has nonzero solutions only provided that its rank is less than $%
2N+2$. Thus we arrive at a symbolic equation $\det ||\mathrm{LHS}|| = 0$.
The determinant of the LHS is an $N$-th order polynomial of $%
(\lambda\kappa)^{-1}$ with complicated (due to the presence of Bessel
functions) coefficients that depend on $\kappa$ and on geometrical
parameters of the system. Zeros of this function with respect to $\kappa$
determine the energy spectrum of the system.

An alternative way to obtain the spectral equation without calculating the
determinant is to eliminate variables. This may be done recursively.
Equations (\ref{systema}, \ref{systemb}) link coefficients $A_{i-1}^{n}$, $%
B_{i-1}^{n}$ to $A_{i}^{n}$, $B_{i}^{n}$ on the next interval. Applying them
successively one can express all $A_{i}^{n}$, $B_{i}^{n}$ in terms of $%
A_{0}^{n}$, $B_{0}^{n}$. At the end, after imposing the boundary conditions (%
\ref{systemc}, \ref{systemd}), we obtain the spectral equation.

We start from equations (\ref{systema}). Let us introduce new variables $%
\Delta _{i}^{n}$, $i=1,\ldots \,N$, associated with the steps of
coefficients $A$ and $B$ at singular points,
\begin{equation}
\Delta
_{i}^{n}=(A_{i-1}^{n}-A_{i}^{n})I_{n}(x_{i})=(B_{i}^{n}-B_{i-1}^{n})K_{n}(x_{i}).
\label{Delta}
\end{equation}
Obviously $A$'s and $B$'s may be expressed in terms of $\Delta $'s:

\begin{mathletters}
\label{A,B,Delta}
\begin{eqnarray}
A_{i}^{n} &=&A_{0}^{n}-\sum_{j=1}^{i}\frac{\Delta _{j}^{n}}{I_{n}(x_{j})};
\label{A,B,Deltaa} \\
B_{i}^{n} &=&B_{0}^{n}+\sum_{j=1}^{i}\frac{\Delta _{j}^{n}}{K_{n}(x_{j})}.
\label{A,B,Deltab}
\end{eqnarray}
\end{mathletters}

On the other hand this representation reduces the number of unknowns by $N$
and guarantees that equations (\ref{systema}) are fulfilled. Note that all $%
A^n_i$ and $B^n_i$ depend only on $\Delta^n_j$ with smaller numbers $j \leq
i $.

Substituting definition (\ref{Delta}) into (\ref{systemb}) we obtain the
following equation for $\Delta $,
\begin{equation}
\Delta _{i}^{n}=\frac{2K_{n}(x_{i})I_{n}(x_{i})\left[
A_{i}^{n}I_{n}(x_{i})+B_{i}^{n}K_{n}(x_{i})\right] }{\kappa \lambda \left[
K_{n}(x_{i})I_{n}^{\prime }(x_{i})-K_{n}^{\prime }(x_{i})I_{n}(x_{i})\right]
}.  \label{Delta-short}
\end{equation}
The expression in square brackets in the denominator is nothing but Wronski
determinant of the modified Bessel functions $I_{n}$ and $K_{n}$
\begin{equation}
W(x)=K_{n}(x)I_{n}^{\prime }(x)-K_{n}^{\prime }(x)I_{n}(x)=\frac{1}{x}.
\label{Wronskian}
\end{equation}
Inserting into (\ref{Delta-short}) formulae (\ref{A,B,Delta}) for $A_{i}^{n}$
and $B_{i}^{n}$ we obtain after obvious cancellations:
\begin{equation}
\Delta _{i}^{n}=\frac{2x_{i}K_{n}(x_{i})I_{n}(x_{i})}{\kappa \lambda }\left[
A_{0}^{n}I_{n}(x_{i})+B_{0}^{n}K_{n}(x_{i})+\sum_{j=1}^{i-1}\Delta
_{j}^{n}\left( \frac{K_{n}(x_{i})}{K_{n}(x_{j})}-\frac{I_{n}(x_{i})}{%
I_{n}(x_{j})}\right) \right] .  \label{Delta-long}
\end{equation}
Thus we managed to express every $\Delta _{i}^{n}$ in terms of the preceding
$\Delta _{j}^{n}$ with $j<i$.

As long as we are interested in the spectrum we may simplify calculations by
taking $A_{0}^{n}=1$, $B_{0}^{n}=0$ in accordance with (\ref{systemc}). (The
correct value of $A_{0}^{n}$ can be found later from the normalization
condition (\ref{normalization}).) Given these values we can find $\Delta
_{1}^{n}$ and then calculate the successive values of $\Delta _{i}^{n}$ $%
A_{i}^{n}$, $B_{i}^{n}$, $i=1,\ldots \,N$. Every new tube brings an extra
power of $(\kappa \lambda )^{-1}$. Thus after $i$ recursions we obtain a
polynomial of power $i$ in $\frac{1}{\kappa \lambda }$.

The last thing is to impose the boundary condition (\ref{systemd}). This
determines the spectrum of the system.

We have already explained that $A^n_N$ is an $N$-th order polynomial in $%
\frac 1{\kappa\lambda}$. However its coefficients contain Bessel functions
that depend on $x_i = \kappa r_i$ and one may not be sure that the equation $%
A^n_N (\kappa,\, \lambda) = 0$ always has $N$ roots in $\kappa$.\footnote{%
We may suggest, though, that for any $\kappa$ there are $N$ values of $%
\lambda_i$ such that $A^n_N (\kappa,\, \lambda_i) = 0$.} Still, as we shall
see later, the roots lie close to each other and variations of Bessel
functions are negligible. Hence in a system of $N$ coaxial nanotubes each
radial energy level is, in general, split into $N$ ones.

\section{The spectral equations}

\subsection{The single tube}

Let us analyze the electronic spectrum for a nanotube of radius $r_t$.
Taking the wave function in form (\ref{psi}) we obtain the equation for
radial part:
\begin{equation}  \label{1tbeqn}
x^2 \psi^{\prime\prime}_n + x \psi^{\prime}_n - (x^2 + n^2) \psi_n = - \frac{%
2 }{\kappa \lambda} x_t^2 \delta(x - x_t)\, \psi_n, \qquad \mathrm{where}
\qquad x_t = \kappa r_t.
\end{equation}
The wave function inside and outside the tube looks as follows:

\begin{mathletters}
\label{1tbpsi}
\begin{eqnarray}
\psi _{n}(x) &=&A_{0}^{n}I_{n}(x)+B_{0}^{n}K_{n}(x)\qquad \mathrm{for}\qquad
x<x_{t};  \label{1tbpsia} \\
\psi _{n}(x) &=&A_{1}^{n}I_{n}(x)+B_{1}^{n}K_{n}(x)\qquad \mathrm{for}\qquad
x>x_{t}.  \label{1tbpsib}
\end{eqnarray}
\end{mathletters}

Coefficients $A$ and $B$ must satisfy equations (\ref{system}) with $N =1$.
As follows from the continuity condition (\ref{systema}) we may introduce an
auxiliary variable $\Delta^n_1$ such that
\begin{equation}  \label{1tbDelta}
\Delta^n_1 = (A^n_0 - A^n_1) I_n(x_t) = (B^n_1 - B^n_0) K_n(x_t).
\end{equation}
For the solitary potential well equation (\ref{Delta-long}) is very simple
and takes the form
\begin{equation}  \label{Delta1}
\Delta^n_1 = \frac{2 x_t K_n(x_t)I_n(x_t)}{\kappa \lambda} \left[ A^n_0
I_n(x_t)+ B^n_0 K_n(x_t) \right ] .
\end{equation}
Now we should impose the boundary conditions and demand the wave function to
be regular at $x = 0$ and $x = \infty$. This means that
\begin{equation}  \label{1tbboundary}
B^n_0 = A^n_1 = 0.
\end{equation}
We have already mentioned that when searching for the spectrum we may assume
that $A^n_0 = 1$. Equating expressions (\ref{1tbDelta}) and (\ref{Delta1})
for $\Delta^n_t$ we arrive at the spectral equation:

\begin{mathletters}
\label{1tbspeqn}
\begin{equation}
\frac{2x_{t}}{\kappa \lambda }K_{n}(x_{t})I_{n}(x_{t})=1  \label{1tbspeqna}
\end{equation}
or in terms of the original physical radius:
\begin{equation}
\frac{2r_{t}}{\lambda }K_{n}(\kappa r_{t})I_{n}(\kappa r_{t})=1
\label{1tbspeqnb}
\end{equation}
\end{mathletters}

Solution of this equation is greatly simplified by the presence of small
parameter $\lambda/r_t \ll 1$. This justifies the use of asymptotic
expressions for the Bessel functions:

\begin{mathletters}
\label{asymptots}
\begin{eqnarray}
I_{n}(x) &\sim &\frac{e^{x}}{\sqrt{2\pi x}}\left[ 1-\frac{1}{2x}(n^{2}-\frac{%
1}{4})+\frac{1}{8x^{2}}(n^{2}-\frac{1}{4})(n^{2}-\frac{9}{4})+o(x^{-2})%
\right] ;  \label{asymptotsa} \\
K_{n}(x) &\sim &\sqrt{\frac{\pi }{2x}}e^{-x}\left[ 1+\frac{1}{2x}(n^{2}-%
\frac{1}{4})+\frac{1}{8x^{2}}(n^{2}-\frac{1}{4})(n^{2}-\frac{9}{4})+o(x^{-2})%
\right] .  \label{asymptotsb}
\end{eqnarray}
\end{mathletters}

Substituting these into (\ref{1tbspeqna}) we get with $(\kappa r_t)^{-2}$
accuracy:
\begin{equation}  \label{1tbkleqn}
1 - \kappa \lambda = \frac 1{2 x_t^2} (n^2 -\frac 14) .
\end{equation}
There must be three roots to this cubic in $\kappa$ equation. Two of them
are small ($\kappa^2 \approx (n^2 -\frac 14)/2 r_t^2$ and lie out of the
domain where asymptotic formulae (\ref{asymptots}) hold. The third one can
be found perturbatively. Taking $\kappa = \kappa^{(0)} + \kappa^{(1)}$ with $%
\kappa^{(1)} \ll \kappa^{(0)} = \lambda^{-1}$ we get for the first
approximation:
\begin{equation}  \label{1tbkappa}
\kappa = \frac 1\lambda \left[1 - \frac {\lambda^2}{2r_t^2} (n^2 -\frac 14) %
\right].
\end{equation}
Note that: a) $\kappa \sim \lambda^{-1}$ is big; b) the correction $%
\lambda(n^2 -\frac 14)/2r_t^2$ is small; c) $x_t \sim r_t/\lambda$ is big
and we may use the asymptotical form (\ref{asymptots}) of Bessel functions.
Thus our approximation is consistent.

According to equation (\ref{kappa}) $\kappa $ determines the electronic
energy spectrum:
\begin{equation}
E_{n}(p_{z})=\frac{p_{z}^{2}}{2m}-\frac{\hbar ^{2}}{2m\lambda ^{2}}+\frac{%
\hbar ^{2}}{2mr_{t}^{2}}\left( n^{2}-\frac{1}{4}\right) =-\frac{mU^{2}}{%
2\hbar ^{2}}-\frac{\hbar ^{2}}{8mr_{t}^{2}}+\frac{p_{z}^{2}}{2m}+\frac{\hbar
^{2}n^{2}}{2mr_{t}^{2}}.  \label{1tbspectrum}
\end{equation}
The origin of all the four addends is transparent. The negative first term $-%
\frac{mU^{2}}{2\hbar ^{2}}$ is the binding energy of electron in the
potential well. It does not depend on the form of graphite layer. The last
two terms are the kinetic energy of the electron: the longitudinal part $%
\frac{p_{z}^{2}}{2m}$ varies continuously whereas the angular component $%
\frac{\hbar ^{2}n^{2}}{2mr_{t}^{2}}$ is quantized. For the tube of radius $%
r_{t}=5\,nm$ the scale of quantization is $\frac{\hbar ^{2}}{2mr_{t}^{2}}%
\approx 1.5\,meV$ . An interesting feature of the spectrum is the
geometrical term $-\frac{\hbar ^{2}}{8mr_{t}^{2}}$ that depends only on the
radius of the tube. Because of it electrons are bound to the tube stronger
than to the plane. The thinner the tube, the stronger the binding. For $%
r_{t}=5\,nm$ it is $\hbar ^{2}/8mr_{t}^{2}\approx 0.38\,meV$).

\subsection{Two tubes}

A new phenomenon that occurs in multiple nanotubes is the electron
tunnelling. In planar systems the hopping of electrons between layers leads
to splitting of degenerate energy levels. Out of the planes the wave
functions of electrons fall exponentially and the splitting is weak compared
to the binding energy. However the latter is large by itself and the effect
might be noticeable. Now we are going to consider the tunnelling in coaxial
nanotubes. Of course the presence of curvature terms ($\propto -\frac{\hbar
^{2}}{mr^{2}}$) eliminates the degeneracy from the very beginning. Still
tunnelling shifts energy levels and enhances their splitting. First we shall
analyze this effect in the system of two coaxial nanotubes.

Let the radii of the tubes be $r_{1}$ and $r_{2}$. The parameter
characteristic of tunnelling is $e^{-\alpha }=\exp \left( -\kappa a\right)
\approx \exp \left( -a/\lambda \right) $, where $a=r_{2}-r_{1}$ is the
interval between the tubes. We shall analyze the electronic spectrum of the
system in the first approximation with respect both to tunnelling and
localization of electrons. Thus we will have at our disposal two small
parameters $e^{-\alpha }$ and $\lambda /r_{1,\,2}$.

Following the recipe of Section~\ref{recursion} we assume that $A^n_0 = 1$, $%
B^n_0 = 0$ and calculate $A^n_2(\kappa,\, \lambda,\, r_1,\, r_2)$. The
boundary conditions require it to be zero. Imposing this condition gives the
spectral equation we are looking for. At first glance the equation $%
A^n_2(\kappa,\, \lambda,\, r_1,\, r_2) = 0$ looks rather clumsy but it
simplifies after neglecting terms of the order $e^{-2\alpha}/x_1$ and
higher:
\begin{equation}  \label{2tbspeqn}
(1 - \lambda\kappa)^2 - \frac 12 \left(n^2 - \frac 14 \right) \left(\frac
1{x_1^2} + \frac 1{x_2^2} \right) (1 - \lambda\kappa) + \frac 1{4 x_1^2
x_2^2} \left(n^2 - \frac 14 \right)^2 - e^{-2\alpha} = 0.
\end{equation}
The coefficients of the equation depend on $\kappa$ via $x^{-2}$ and $%
e^{-2\alpha}$. However this dependence is weak. We shall consider them as
constants and substitute for $\kappa$ the zeroth approximation value $%
\kappa^{(0)} = \lambda^{-1}$. Then the equation becomes a simple quadratic
one. The two solutions to it are
\begin{equation}  \label{2tbkleqn}
(1 - \lambda\kappa)_{1,\, 2} = \frac 14 \left(n^2 - \frac 14 \right)
\left(\frac 1{x_1^2} + \frac 1{x_2^2} \right) \pm \sqrt{\frac 1{16}
\left(n^2 - \frac 14 \right)^2 \left(\frac 1{x_1^2} - \frac 1{x_2^2}
\right)^2 + e^{-2\alpha}}.
\end{equation}

Calculating the energy levels from equation (\ref{kappa}) we get:
\begin{eqnarray}
\lefteqn{E_n(p_z)_{1,\, 2} = - \frac{\hbar^2}{2m\lambda^2} + \frac{p^2_z}{2m}
} \hspace{10mm} &&  \nonumber \\
& + & \frac{\hbar^2}{4m} \left[ \left(n^2-\frac 14 \right) \left(\frac
1{r_1^2} + \frac 1{r_2^2} \right) \mp \sqrt{\left(n^2 - \frac 14 \right)^2
\left(\frac 1{r_1^2} - \frac 1{r_2^2} \right)^2 + \frac{16 e^{-2\alpha}}{%
\lambda^4}}\right].  \label{2tbspectrum}
\end{eqnarray}

Let us discuss the limiting cases. Suppose that the attractive potential be
strong enough to prevent tunnelling. This corresponds to $e^{-2\alpha}
\rightarrow 0$ and only $r^{-2}$ terms (\emph{i.~e.\/} the rotational part
of kinetic energy) should be taken into account. The two roots of equation (%
\ref{2tbkleqn}) are

\begin{mathletters}
\label{2tbnotunnel}
\begin{equation}
(1-\lambda \kappa )_{1,\,2}=\frac{1}{2x_{1,\,2}^{2}}\left( n^{2}-\frac{1}{4}%
\right) \qquad \mathrm{for}\qquad 16e^{-2\alpha }\ll \left( n^{2}-\frac{1}{4}%
\right) ^{2}\left( \frac{1}{x_{1}^{2}}-\frac{1}{x_{2}^{2}}\right) ^{2}.
\label{2tbnotunnela}
\end{equation}
That corresponds to the energies
\begin{equation}
E_{n}(p_{z})_{1,\,2}=-\frac{\hbar ^{2}}{2m\lambda ^{2}}+\frac{p_{z}^{2}}{2m}+%
\frac{\hbar ^{2}\left( n^{2}-\frac{1}{4}\right) }{2mr_{1,\,2}^{2}}\quad
\mathrm{for}\quad \frac{16e^{-\frac{2a}{\lambda }}}{\lambda ^{4}}\ll \left(
n^{2}-\frac{1}{4}\right) ^{2}\left( \frac{1}{r_{1}^{2}}-\frac{1}{r_{2}^{2}}%
\right) ^{2}.  \label{2tbnotunnelb}
\end{equation}
\end{mathletters}

Comparing this with (\ref{1tbkleqn}) we see that the spectrum consists of
two independent series coming from the noninteracting tubes of radii $%
r_{1,\, 2}$. This is exactly what one should expect in absence of tunnelling.

The second case corresponds to large radii and small distance between the
tubes. Now we neglect $x_{1,\, 2}^{-2} \ll e^{-2\alpha}$. Now the solutions
become

\begin{mathletters}
\label{2tbflat}
\begin{equation}
(1-\lambda \kappa )_{1,\,2}=\pm \exp \left( -\alpha \right) \qquad \mathrm{%
for}\qquad 16e^{-2\alpha }\gg \left( n^{2}-\frac{1}{4}\right) ^{2}\left(
\frac{1}{x_{1}^{2}}-\frac{1}{x_{2}^{2}}\right) ^{2},  \label{2tbflata}
\end{equation}
and the energies are
\begin{equation}
E_{n}(p_{z})_{1,\,2}=-\frac{\hbar ^{2}}{2m\lambda ^{2}}\left( 1\mp 2\exp
\left( -\frac{a}{\lambda }\right) \right) +\frac{p_{z}^{2}}{2m}\quad \mathrm{%
for}\quad \frac{16e^{-\frac{2a}{\lambda }}}{\lambda ^{4}}\gg \left( n^{2}-%
\frac{1}{4}\right) ^{2}\left( \frac{1}{r_{1}^{2}}-\frac{1}{r_{2}^{2}}\right)
^{2}.  \label{2tbflatb}
\end{equation}
\end{mathletters}

This coincides with the splitting of the ground level that arises in two
parallel planes due to electron tunnelling between them. Note, however, that
because of the factor $\left( n^{2}-\frac{1}{4}\right) $ dropping down $%
x^{-2}$ terms is legitimate only for low-lying levels whereas for large $n$
the kinetic part of (\ref{2tbspectrum}) always comes into play. When neither
of the terms under the square root is small the two effects add up. Thus we
may conclude that tunnelling enhances the splitting of equal $n$ energy
levels in coaxial systems. This effect is more relevant for small $n$ and
becomes less for upper levels.

Summarizing this Section we notice, that the value of tunnel splitting $%
\Delta E_{tunnel}=\frac{2\hbar ^{2}}{m\lambda ^{2}}e^{-\frac{a}{\lambda }}$
depends on distance between the tubes. It reaches the maximum for $%
a=2\lambda $:
\begin{equation}
\Delta E_{tunnel}^{max}=\frac{2\hbar ^{2}}{m\lambda ^{2}}e^{-2}\approx 0.27%
\frac{\hbar ^{2}}{m\lambda ^{2}}.  \label{2tbtunmax}
\end{equation}
In real systems $\lambda $ is on atomic scales (several angstroms) and $a$
is several nanometers. Therefore probably the maximum is inaccessible. In
order to increase the role of tunnelling one has to make the tubes closer to
each other.

We can also estimate the number $n$ starting from which the effect of
tunnelling becomes less important and the systems behaves like the set of
independent tubes. In order to obtain the upper bound let us assume that $%
a=2\lambda \ll r_{1,\,2}$. Then inequality (\ref{2tbnotunnelb}) turns into
\begin{equation}
n>\sqrt{\frac{2}{e}}\left( \frac{r}{\lambda }\right) ^{\frac{3}{2}%
}=0.86\left( \frac{r}{\lambda }\right) ^{\frac{3}{2}}.  \label{nmin}
\end{equation}
This is a large value. Certainly in real systems with $a\gg \lambda $ the
tubes become independent much earlier. Actually for $n$ this big the kinetic
part in (\ref{2tbnotunnelb}) exceeds the binding energy. Therefore the total
energy is positive and the entire model fails. An improvement of this
estimate requires a better knowledge of $\lambda $.

We see that the tunnelling of electrons between coaxial nanotubes enhances
the splitting of energy levels with equal $n$. This effect is stronger
expressed for low-lying levels. For big $n$ the role of tunnelling falls
down and the energy spectrum approaches the superposition of those of
several noninteracting tubes. The crucial parameter for tunnelling is $\exp
\left( -a/\lambda \right) $. In our model $\lambda $ is a function of the
energy of electron binding to conducting layers. The tunnel interaction
enhances as the tubes get closer and the interval between them becomes
small. An estimate for the tunnelling parameter may be obtained from the
probability of electron ``hopping'' between the layers in planar systems.

\section{Physical properties}

\subsection{$\protect\bigskip $The density of states}

As it is well known the one-particle density of states (DOS) can be
expressed in terms of the imaginary part of one-electron Green function. In
the model of MWCN introduced above the quasiparticle spectrum has a form:

\begin{equation}
\xi (p_{z},n,k)=\frac{p_{z}^{2}}{2m}+\Delta E(n,k)-E_{F},
\end{equation}
where $\Delta E(n,k)$ denotes the energy related to circumferential modes
quantization and splitting due to the inter-tube tunnelling. It may be
numerically calculated for MWNT with any number of conducting cylinders with
the help of the formalism developed in Secs. 2 and 3. Index $n$ enumerates
the levels of angular quantization, while index $k$ enumerates the energy
levels related to electron hopping between adjacent tubes (fine structure of
energy levels).

The electron Green function is, therefore,

\begin{equation}
G^{R}(p_{z},n,k;\epsilon )=\frac{1}{\epsilon -\frac{p_{z}^{2}}{2m}-\Delta
E(n,k)+E_{F}+\frac{i}{2\tau }},
\end{equation}
($\tau $ is the time of one-electron elastic relaxation).

The DOS of electron in the multi-wall nanotube is now given by

\begin{equation}
\nu (\epsilon )=-\frac{2}{\pi }\mbox{Im}\int_{-\infty }^{\infty }\frac{dp_{z}%
}{2\pi }\sum_{n,k}\frac{1}{\epsilon -\frac{^{p_{z}^{2}}}{2m}-\Delta
E(n,k)+E_{F}+\frac{i}{2\tau }}.
\end{equation}
The summation over $n$ should be performed to take into account all the
occupied electronic states, \emph{i.e.} from $-N_{\max }$ to $N_{\max }$,
where $N_{\max }$ is determined by the value of electron chemical potential.
A typical value for a realistic nanotube is $N_{\max }\thickapprox 5$\cite
{EG01}. The summation over $k$ should involve all energy levels of the fine
structure. The numbers of those is equal to the number of coaxial tubes in
MWNT. After integration over $p_{z}$ and straightforward algebra we get:

\begin{equation}
\nu (\epsilon )=\frac{\sqrt{2m}}{\pi }\sum_{n,k}\frac{1}{\sqrt{\left[
E_{F}-\Delta E(n,k)+\epsilon \right] }}.
\end{equation}

In order to calculate thermodynamic functions in terms of Green function we
follow the formalism of Ref. \cite{AGD}. Inasmuch as the analytical
properties of the Green function are the same as in the case of normal 3D
metal, we may use the result of Ref. \cite{AGD} after summation over $%
\epsilon $. In the case of MWNT we have:

\begin{eqnarray}
\frac{S}{V} &=&\frac{2\pi T}{3}\int_{-\infty }^{\infty }\frac{dp_{z}}{2\pi }%
\sum_{n,k}\mbox{Im}\left[ \frac{1}{G^{R}(p_{z},n,k;\epsilon )}\frac{\partial
G^{R}(p_{z},n,k;\epsilon )}{\partial \epsilon }\right] _{\epsilon =0} \\
&=&\frac{2\pi T}{3}\mbox{Im}\int_{-\infty }^{\infty }\frac{dp_{z}}{2\pi }%
\sum_{n,k}\frac{1}{\frac{p_{z}^{2}}{2m}+\Delta E(n,k)-E_{F}-\frac{i}{2\tau }}%
,  \nonumber
\end{eqnarray}
\bigskip and we see that the entropy is proportional to DOS:

\begin{equation}
\frac{S}{V}=\frac{4\pi }{3}T\nu (0).
\end{equation}

\subsection{The tunnel current}

$\bigskip $We study now the tunnel current between the nanotube and normal
metallic contact. The general expression for the tunnel current is
\begin{equation}
I(V)=-e\mbox{Im}\left[ K^{R}(\omega _{\nu })\right] _{i\omega _{\nu }=eV},
\end{equation}
where

\begin{equation}
K^{R}(\omega _{\nu })=T\sum_{\varepsilon _{m}}\int_{-\infty }^{\infty }\frac{%
dp_{z}}{2\pi }\sum_{n,k}\int \frac{d^{3}\mathbf{k}}{\left( 2\pi \right) ^{3}}%
\left| T_{\mathbf{p},\mathbf{k}}\right| ^{2}G(p_{z},n,k;\varepsilon
_{m}+\omega _{\nu })G^{(0)}(\mathbf{k},\varepsilon _{m}).  \nonumber
\end{equation}
Here the superscript $(0)$ refers to the normal metal. Integration of  the
corresponding Green function over momenta gives
\begin{equation}
\int \frac{d^{3}\mathbf{k}}{\left( 2\pi \right) ^{3}}G^{(0)}(\mathbf{k}%
,\varepsilon _{m})=-i\pi \nu ^{(0)}\mathop{\mathrm{sgn}}(\varepsilon _{m}).
\end{equation}
One can also easily perform the  $p_{z}$ integration of the Green function $%
G(p_{z},n,k;\varepsilon _{m}+\omega _{\nu })$:

\begin{equation}
\int_{-\infty }^{\infty }\frac{dp_{z}}{2\pi }G(p_{z},n,k;\varepsilon
_{m}+\omega _{\nu })=-\frac{i\pi \sqrt{m}\mathop{\mathrm{sgn}}(\varepsilon
_{m}+\omega _{\nu })}{\sqrt{2\left[ E_{F}-\Delta E(n,k)+i\varepsilon _{m+\nu
}\right] }}.
\end{equation}

The calculation of the sum over $\varepsilon _{m}$ is straightforward:

\begin{eqnarray}
&&\mbox{Im}\left[ T\sum_{\varepsilon _{m}}\frac{\mathop{\mathrm{sgn}}%
(\varepsilon _{m})\mathop{\mathrm{sgn}}(\varepsilon _{m}+\omega _{\nu })}{%
\sqrt{-\Delta E(n,k)+E_{F}+i\varepsilon _{m+\nu }}}\right] _{i\omega _{\nu
}=eV}= \\
&&\frac{1}{2\pi }\int_{-\infty }^{\infty }d\epsilon \left( \tanh \frac{%
\epsilon +eV}{2T}-\tanh \frac{\epsilon }{2T}\right) \frac{1}{\sqrt{%
E_{F}-\Delta E(n,k)+\epsilon }}.  \nonumber
\end{eqnarray}
Finally, for the case of low voltage we can expand in $eV$ and obtain the
Ohm's law:

\begin{equation}
I(V)=2e^{2}V\nu ^{(0)}(0)\nu (0)\left\langle \left| T_{\mathbf{p},\mathbf{k}%
}\right| ^{2}\right\rangle .
\end{equation}
For intermediate voltages the non-linear regime in tunnel conductance may be
evaluated numerically:
\begin{equation}
G_{tun}\left( V\right) =\frac{dI(V)}{dV}=e^{2}\nu ^{(0)}\sqrt{m}\frac{\pi }{%
4T}\left\langle \left| T_{\mathbf{p},\mathbf{k}}\right| ^{2}\right\rangle
\sum_{n,k}\int_{-\infty }^{\infty }\frac{d\epsilon }{\cosh ^{2}\frac{%
\epsilon }{2T}}\frac{1}{\sqrt{\epsilon +E_{F}-\Delta E(n,k)-eV}}.
\label{sigmatunn}
\end{equation}
When the temperature is not too low ($E_{F}-\Delta E(n_{\max },k_{\max })\ll
T\ll \frac{\hbar ^{2}}{2mr_{t}^{2}}$) one can substitute the $\cosh ^{2}%
\frac{\epsilon }{2T}$ by unity and using the fact that the distance between
the levels of angular quantization is still much larger than the temperature
one gets
\begin{equation}
G_{tun}\left( V\right) \approx e^{2}N_{tub}\nu ^{(0)}\sqrt{m}\frac{\pi }{2T}%
\left\langle \left| T_{\mathbf{p},\mathbf{k}}\right| ^{2}\right\rangle \sqrt{%
2T-eV}.
\end{equation}
In the limit of very low temperatures $T\ll E_{F}-\Delta E(n_{\max },k_{\max
})$ the $\cosh ^{-2}\frac{\epsilon }{2T}$ works as the delta-function and
the tunnelling conductance reproduces the energy dependence of density of
states:
\begin{equation}
G_{tun}\left( V\right) =\pi e^{2}\nu ^{(0)}\sqrt{m}\left\langle \left| T_{%
\mathbf{p},\mathbf{k}}\right| ^{2}\right\rangle \sum_{n,k}\frac{1}{\sqrt{%
E_{F}-\Delta E(n,k)-eV}}=\frac{\pi ^{2}e^{2}\nu ^{(0)}}{\sqrt{2}}%
\left\langle \left| T_{\mathbf{p},\mathbf{k}}\right| ^{2}\right\rangle \nu
\left( -eV\right) .
\end{equation}
This result is valid for temperatures less than the fine energy-level
splitting $\ T\ll \frac{\hbar ^{2}}{2m\lambda ^{2}}\exp \left( -\frac{a}{%
\lambda }\right) \sim 1K$.

\section{Summary}

Let us summarize the results. Using the Kronig-Penney type potential we
model the MWNT as a cylindrical crystal. Because of the circumferential
modes quantization, the electron states in the individual tube form a
sequence of one-dimensional sub-bands. A new phenomenon that occurs in
multiple nanotubes is the electron tunnelling between the adjacent tubes.
The hopping of electrons between the tubes leads to splitting of otherwise
degenerate energy levels. We have developed the general formalism which
allows to evaluate numerically the energy levels of nanotube composed of any
number of conducting cylinders. We have demonstrated that tunnelling
enhances the splitting of energy levels with equal $n$ in coaxial systems.
This effect is more relevant for small $n$ and becomes less for upper
levels. The crucial parameter for tunnelling is $\exp \left( -a/\lambda
\right) ,$ where $a$ is the distance between the tubes and $\lambda $ is the
effective width of $\delta $-potential. Naturally, the tunnel interaction
enhances as the tubes get closer and the interval between them becomes small.

We applied the Green function formalism and calculated the electron
spectrum, the one-particle density of states and entropy. Moreover, we
demonstrated the possibility of direct experimental measurement of DOS
structure by means of the tunnelling microscopy. The calculated within our
approach theoretical dependence of tunnel conductance on voltage is
presented for realistic set of physical parameters in Fig. 1. In this plot
we choose the parameters typical for MWNT: $U=1$ $Ry$ (pseudopotential of
carbon), $a=0.3$ $nm$, and $R=17$ $nm$. Notice, that in this case the
effective width of attractive singular potential $\lambda $ is $0.02$ $nm$
confirming our approximation of strong potential $U\gg \frac{\hbar ^{2}}{%
ma^{2}}$. One can see the resemblance between the fine structure of peaks of
the theoretical density of states and experimentally observed peaks of
tunnelling conductance \cite{ZCDR03}.

An important issue is the possibility of direct experimental studies of DOS
by methods of tunnelling spectroscopy. According to Eq. (\ref{sigmatunn}),
the tunnelling conductance is the convolution of two functions depending on
the electron energy. This equation results in direct proportionality between
tunnelling conductance and DOS in the limiting case of very low temperature.
Namely, the temperature should be much less than the energy scale of DOS to
be resolved. Therefore the interpretation of results of tunnelling
spectroscopy needs care. In particular, for resolving the fine structure of
DOS due to the intertube hopping the temperature should be not higher that
few tenth of Kelvin.

Our theory is based on the analysis of the non-interacting electron system
and does not explain the appearance of the experimentally observed
low-temperature zero bias anomaly in MWNT \cite{WVRS98,SBSS99}.
Nevertheless, it is clear that such dip in the tunnelling conductance can be
attributed to the effects of electron-electron interaction either in
diffusion channel (Altshuler-Aronov-Lee effect) \cite{AAL80,AA85} or in
Cooper channel (superconducting fluctuations) \cite{VD83}. In the first
scenario the width of the dip on the voltage scale turns out to be of the
order of $\tau ^{-1}$($\tau $ is elastic scattering time), while in the
second one it is of the order of $\ \pi T.$ For the case when zero-bias
anomaly is large in magnitude, the nonperturbative approaches have to be
applied \cite{EG01}. Anyway, experimental data should reflect both free- and
interacting electron effects in tunnel resistance. This would lead to a
complicated dependence on voltage as it was really observed in MWNT.

Another important aspect of direct probing of intrinsic electronic
properties of carbon nanotubes concerns the experimental problem of making
low-resistance contacts to the nanotube. Indeed, if the resistance of the
contact between a nanotube and metallic electrode is much higher than the
resistance of the nanotube itself, the low-temperature data would refer to
the properties of the metal-nanotube contact, rather than to intrinsic
properties of the nanotube \cite{ZCDR03,KKFD03}.

Authors acknowledge financial support from the Program ``New Materials'',
``Universities of Russia'' of Russian Ministry of Education and RFBR grant
03--02--16209 , Italian projects FIRB ``TIN'', COFIN 2004.

\newpage
\begin{figure}[tbp]
\centerline{\includegraphics[width=.8\textwidth]{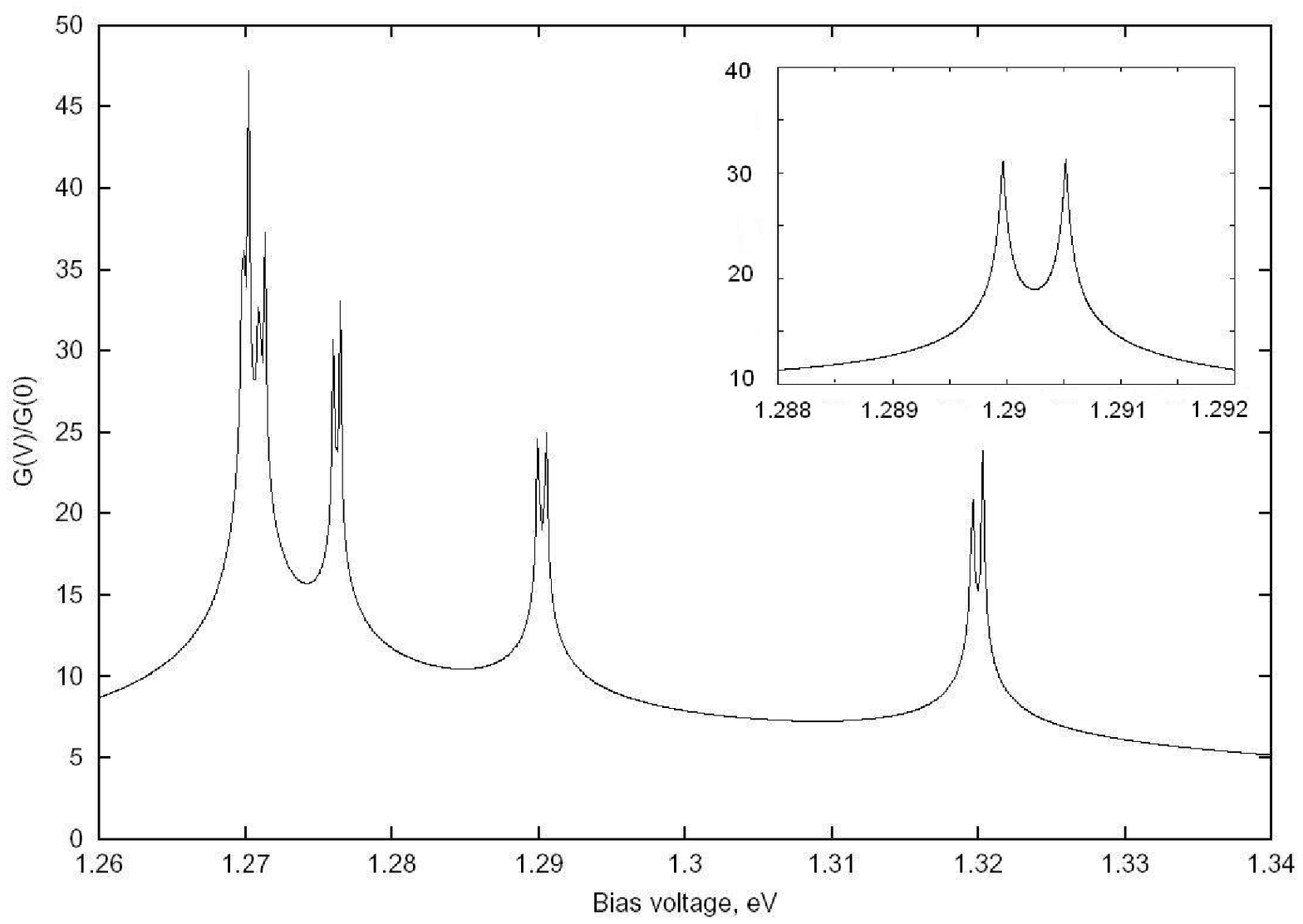}} \caption{ The
theoretical dependence of tunnel conductance on voltage for realistic set of
physical parameters.}\label{Fig.1}
\end{figure}
\end{document}